\documentclass[conference]{IEEEtran}
\IEEEoverridecommandlockouts
\usepackage{cite}
\usepackage{amsmath,amssymb,amsfonts}
\usepackage{algorithmic}
\usepackage[pdftex]{graphicx}
\usepackage{textcomp}
\usepackage{xcolor}
\def\BibTeX{{\rm B\kern-.05em{\sc i\kern-.025em b}\kern-.08em
    T\kern-.1667em\lower.7ex\hbox{E}\kern-.125emX}}

\makeatletter
\def\endthebibliography{%
    \def\@noitemerr{\@latex@warning{Empty `thebibliography' environment}}%
    \endlist
}
\makeatother

\begin{document}

\title{Wireless Link Quality Estimation \\ Using LSTM Model\\
}

\author{\IEEEauthorblockN{1\textsuperscript{st} Yuki Kanto}
\IEEEauthorblockA{\textit{Graduate School of Engineering} \\
\textit{Nagaoka University of Technology}\\
Nagaoka, Niigata, Japan \\
s213121@stn.nagaokaut.ac.jp}
\and
\IEEEauthorblockN{2\textsuperscript{nd} Kohei Watabe}
\IEEEauthorblockA{\textit{Graduate School of Engineering} \\
\textit{Nagaoka University of Technology}\\
Nagaoka, Niigata, Japan \\
k\_watabe@vos.nagaokaut.ac.jp}
}

\maketitle

\begin{abstract}
    In recent years, various services have been provided through high-speed and high-capacity wireless networks on mobile communication devices, necessitating stable communication regardless of indoor or outdoor environments.
    To achieve stable communication, it is essential to implement proactive measures, such as switching to an alternative path and ensuring data buffering before the communication quality becomes unstable.
    The technology of Wireless Link Quality Estimation (WLQE), which predicts the communication quality of wireless networks in advance, plays a crucial role in this context.
    In this paper, we propose a novel WLQE model for estimating the communication quality of wireless networks by leveraging sequential information. 
    Our proposed method is based on Long Short-Term Memory (LSTM), enabling highly accurate estimation by considering the sequential information of link quality.
    We conducted a comparative evaluation with the conventional model, stacked autoencoder-based link quality estimator (LQE-SAE), using a dataset recorded in real-world environmental conditions. 
    Our LSTM-based LQE model demonstrates its superiority, achieving a 4.0\% higher accuracy and a 4.6\% higher macro-F1 score than the LQE-SAE model in the evaluation. 
\end{abstract}


\section{Introduction}
In recent years, mobile communication devices have facilitated a variety of services through high-speed, high-capacity wireless networks. Ensuring stable communication in both indoor and outdoor environments has become a critical requirement for these services. 
These mobile communication devices encompass a wide range of devices, including smartphones, autonomous mobility systems, Unmanned Aerial Vehicles (UAVs), IoT devices, and more. Depending on the nature of the services provided, network connectivity is sometimes indispensable for operational purposes.
For services that are highly dependent on wireless network connectivity, the stability and reliability of link quality are the most crucial factors in maintaining service quality.

To provide a stable communication environment, effective measures, such as preemptively switching to optimal communication paths and buffering data before instability occurs, are essential. 
Therefore, the technology of Wireless Link Quality Estimation (WLQE) is vital for estimating the communication quality of current and future wireless networks.
WLQE enables the quantitative evaluation of wireless network quality, facilitating real-time or anticipatory optimization of networks. 
This, in turn, ensures stable service quality through mobile devices, resulting in a good user experience and safe device operation.

In recent years, a shift in WLQE research to machine learning approaches using link quality data has been observed; however, the significant potential of using sequential data in estimating link quality has largely been overlooked.  
Numerous WLQE methods using machine learning have been proposed, including approaches using random forests, logistic regression, and Artificial Neural Networks (ANN)~\cite{cerar2021survey}. 
The stacked autoencoder-based link quality estimator (LQE-SAE)~\cite{luo2019link} that uses deep learning is a state-of-the-art WLQE method. However, it does not use sequential data as input, and therefore, it cannot fully exploit the characteristics of link quality that dynamically change over time.

In this paper, we propose an LSTM-based LQE model that leverages sequential information using Long Short-Term Memory (LSTM) to accurately estimate future link quality. 
The LSTM-based LQE model processes metrics related to past link quality as sequential data using LSTM and outputs the grade of future link quality determined by binning. 
LSTM allows for the better capture of latent patterns and temporal dependencies within a feature space of sequence data, leading to higher classification accuracy of the LQE grades. 
The performance of the proposed method is evaluated by comparing it with the conventional LQE-SAE using a dataset recorded link quality in real-world environments.

The rest of this paper is structured as follows.
Section~\ref{sec:RelatedWorks} describes related research in the WLQE field.
The proposed method is detailed in Section~\ref{sec:ProposedMethod}.
Following that, Section~\ref{sec:ExperimentalResults} addresses the evaluation of the proposed and conventional methods' performance, discussing evaluation methods and results.
In Section~\ref{sec:Conclusion}, we present the conclusion of this study and future challenges.

\section{Relted Works}\label{sec:RelatedWorks}
Research in the field of WLQE began in the late 1990s, primarily focusing on statistical approaches~\cite{cerar2021survey}. Early LQE models indicated that wireless transmission methods are prone to significant packet loss compared with wired transmission methods. In 1996, a statistical error model based on the distribution of packet stream errors and error-free lengths improved TCP throughput~\cite{nguyen1996trace}.
Research on LQE has led to various models, including statistical models such as Markov models~\cite{nguyen1996trace,6gc4-y070-22,balakrishnan1998explicit,woo2003taming,sing-08-01}, rule-based models, and threshold-based models~\cite{4blehotnets2007,boano2010triangle,audeoud2018quick}. In addition, models that combine fuzzy logic and machine learning, known as fuzzy ML models, have also been proposed~\cite{baccour2010f,guo2013fuzzy,rekik2015low,baccour2011radiale}.

Since 2010, LQE research has shifted toward utilizing machine learning techniques such as naive bayes classifiers, logistic regression, and ANN to estimate Packet Reception Ratio (PRR) based on link quality metrics like PRR, Received Signal Strength Indicator (RSSI), and Signal-Noise Ratio (SNR)~\cite{liu2011foresee}.. Following these developments, wireless link quality estimation models using various machine learning techniques have gained attention.
Notable machine learning approaches include statistical ML models~\cite{demetri2019automated,liu2011foresee,liu2014temporal,liu2012talent,cerar2020designing,millan2015time,okamoto2017machine,bote2018online,shu2017research,rehan2016machine}, reinforcement learning models~\cite{ancillotti2017reinforcement}, and deep learning models~\cite{sun2017wnn,luo2019link}. 
In particular, LQE-SAE, categorized under deep learning models, can accurately estimate link quality grades by inputting three features (i.e., RSSI, SNR, and LQI) into the Stacked AutoEncoder (SAE)~\cite{luo2019link}. It is considered to be one of the most superior LQE methods.
The shift toward machine learning approaches can be attributed to the limitations of statistical methods, which rely on modeling with limited datasets and features. These limitations hindered adaptability to real-world environments and estimation accuracy. 
Consequently, recent research has focused on efficiently learning from large datasets and multiple features, including SNR and RSSI, to develop highly accurate and adaptable machine learning-based models.

\section{Proposed Method}\label{sec:ProposedMethod}
We propose a novel LQE method that enables highly accurate link quality estimation using an LSTM-based model. We consider that conventional machine learning-based LQE methods do not effectively leverage sequential information.
Our proposed method uses LSTM to predict the Reference Signal Received Power (RSRP) metric as a regression model. Subsequently, the predicted link quality is classified into five grades (very good, good, intermediate, bad, and very bad) using a multi-class classification approach based on binning. 
LSTM takes a sequence of link quality metrics, including RSRP, arranged over time as input features. The output represents the link quality grade, determined by binning, of the RSRP value predicted by LSTM for future time points.

\subsection{Preprocess}\label{ssec:Preprocess}
In our proposed method, we preprocess the input data before feeding a dataset of continuously observed link quality metrics, including RSRP, as sequence data. 
We performed five preprocessing steps: missing value imputation, oversampling, noise separation, standardization, and binning.

For missing value imputation and oversampling, we adopt commonly used techniques in the field of machine learning.
For missing value imputation in the dataset, we employ zero padding.
Additionally, to balance the class distribution of link quality grades, we apply Random Over Sampling (ROS) to the minority class data to match the number of samples in the majority class.

Since link quality metrics such as RSRP often exhibit significant variability and contain noisy data, we separate them into a long-term trend term and a noise term using Exponential Moving Average (EMA). The trend term $\Tilde{x}^j_t$ for the sequence $x^j_t;(t=0, 1, 2, \dots)$ of the $j$-th feature is defined as follows:
\begin{align}\label{eq:EMA}
    \Tilde{x}^j_0 &= x^j_{0}, \\
    \Tilde{x}^j_t &= \alpha x^j_{t} + (1 - \alpha)\Tilde{x}^j_{t-1},
\end{align}
where, $\alpha$ is the smoothing factor, defined by the span coefficient $\tau$ as:
\begin{equation}\label{eq:EMA-alpha}
    \alpha = \frac{2}{\tau+1.0}. 
\end{equation}
The noise term $\varepsilon^j_t$ is the difference between the actual value $x^j_t$ and the trend term $\Tilde{x}^j_t$:
\begin{equation}\label{eq:noise}
    \varepsilon^j_t = x^j_t - \Tilde{x}^j_t. 
\end{equation}

To ensure that each feature input has the same scale, we perform standardization by setting the mean to zero and the variance to one for all features. This is done because LQI matrices have different units and tend to vary widely.

The binning method used to convert the continuous RSRP estimation values output by LSTM into multi-class link quality grades is based on the bin widths for each link quality category, as shown in TABLE~\ref{tab:rsrp}.

\begin{table}[tbp]
    \centering
    \caption{Mapping of RSRP Values to Link Quality}
    \begin{tabular}{c|c}
        RSRP{[}dBm{]} & Link Quality \\\hline\hline
        $-84 \leq$ RSRP & Very Good \\
        $-84 <$ RSRP $< -95$ & Good \\
        $-95 <$ RSRP $< -105$ & Intermediate \\
        $-105 <$ RSRP $< -115$ & Bad \\
        RSRP $\leq -115$ & Very Bad \\
    \end{tabular}
    \label{tab:rsrp}
\end{table}

\subsection{Model Architecture}\label{ssec:ModelArchitecture}
\begin{figure}[tbp]
    \centering
    \includegraphics[width=3.5in]{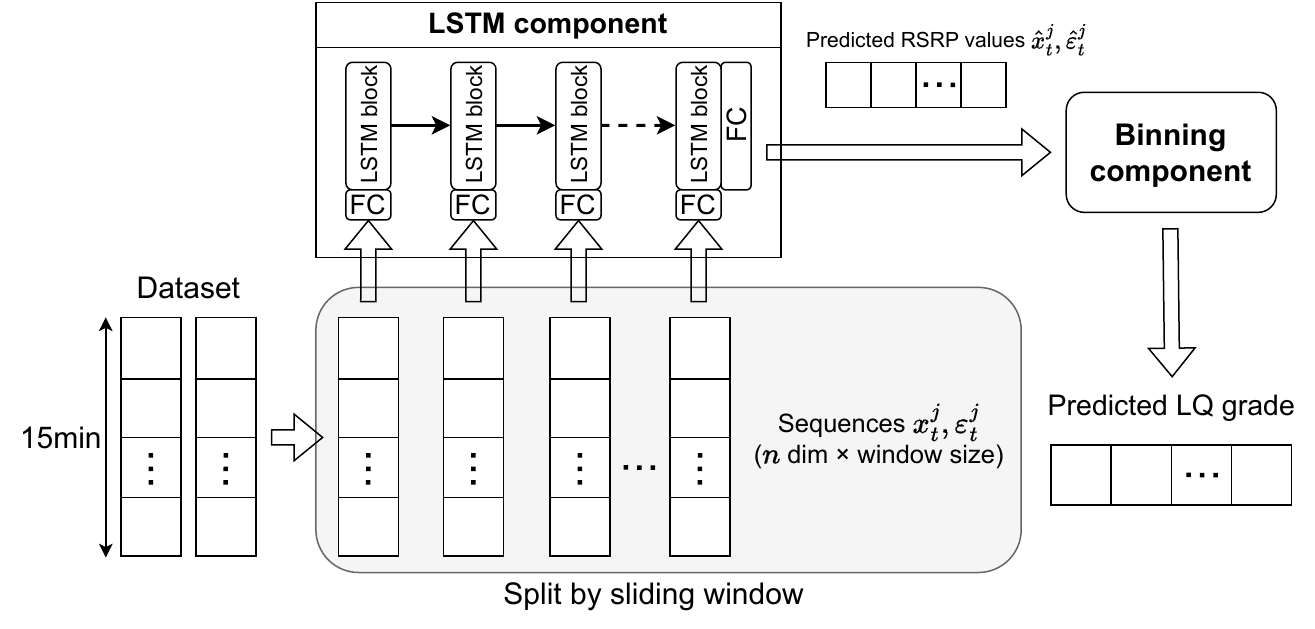}
    \caption{The architecture of LSTM-based LQE model. 
    This model first divides the dataset using a sliding window. Subsequently, two sequences of the trend term $x^j_t$ and the noise term ${\varepsilon}^j_t$ are created from the divided sequence. 
    These sequences are then fed into the LSTM component, predicting the future values of the two terms. The binning component classify it into an LQ grade.}
    \label{fig:architecture}
\end{figure}

The architecture of our proposed method consists of two components, as illustrated in Fig.~\ref{fig:architecture}: the LSTM component and the binning component.
The LSTM component learns short sequence data of length $N$ that are split from long sequence data in a dataset by a sliding window. The short sequence is a series of vectors whose elements represent the trend and noise terms of $n$ features, including RSRP. 
This sequence is represented as $[\tilde{x}^1_t, \tilde{x}^2_t, \dots, \tilde{x}^n_t, \tilde{\varepsilon}^1_t, \tilde{\varepsilon}^2_t, \dots \tilde{\varepsilon}^n_t]^\mathrm{T}\;(t=0, 1, 2, \dots , N-1)$. The output of the LSTM component is the RSRP values $\tilde{x}^1_N$ and $\tilde{\varepsilon}^1_N$ one time slot ahead, serving as a regression model to predict future RSRP values.
After the output from the LSTM component, we obtain the predicted value $\tilde{x}^1_N$ of RSRP by adding $\tilde{x}^1_N$ and $\tilde{\varepsilon}^1_N$ output by the LSTM component.
Note that the first feature $x^1_t$ represents the RSRP that we intend to estimate.

On the other hand, the binning component classifies the link quality grade using a binning method that determines the continuous RSRP estimation values output by LSTM based on bin widths. LSTM models are effective for processing sequential data, but they are not well-suited for estimating categorical link quality grades. 
Therefore, the binning method is employed for multi-class classification to estimate the link quality grade.

\subsection{Training}\label{ssec:TrainingAlgorithm}
In our proposed method, we optimize the parameters of the LSTM component by training it with the dataset.
The training of the LSTM component begins by preprocessing the dataset using the method described in Section~\ref{ssec:Preprocess}. 
The dataset is then divided into training and validation datasets. Subsequently, the divided dataset is used to create input data sequences of length equal to the window size $N$ using a sliding window approach. 
Label data are created from the RSRP values of the next time slot of these input data.
The training progresses on the training dataset, aiming to minimize the value of the loss function on the validation dataset during the learning process. Early stopping is implemented when there is no further improvement in the loss function value, aiming to shorten the learning time by terminating the training process earlier.
After early stopping, the trained model is selected as the model with the minimum loss function value on the validation dataset.

The loss function is defined as the Mean Squared Error (MSE) between the estimated RSRP values $\hat{x}_i$ for the $i$-th input data and the corresponding label data $x_i$. 
The definition of the loss function is as follows: 
\begin{equation}\label{eq:MSE}
    \mathrm{MSE} = \frac{1}{n}\sum_{i=1}^{n}(\hat{x}_i - x_i)^{2}. 
\end{equation}

\section{Experiments}\label{sec:ExperimentalResults}
We conducted training and estimation using the proposed LSTM-based LQE model with a dataset recorded from actual mobile communication devices. 
To compare with the conventional LQE-SAE model, we used the same dataset for training and prediction of the LQE-SAE model, and compared the models based on accuracy and F1 score.
Through this evaluation, we demonstrate that the LSTM-based LQE model outperforms the conventional LQE-SAE model in terms of both accuracy and F1 score.
 
\subsection{Dataset}\label{ssec:Dataset}
\begin{figure}[tbp]
    \centering
    \includegraphics[width=3.5in]{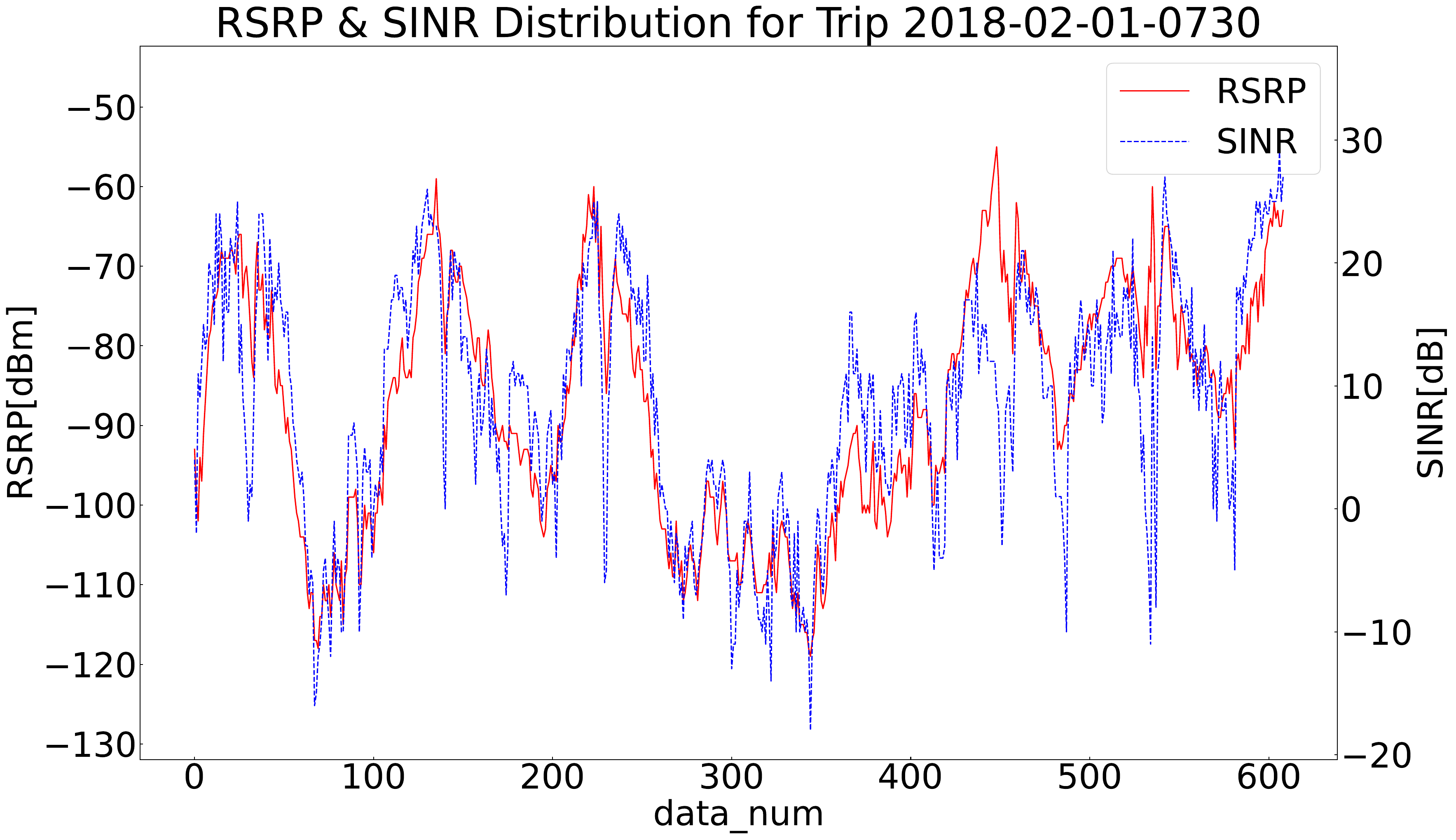}
    \caption{Temporal variation in RSSI and SINR samples in the SRFG dataset. The average RSRP for this sample is $-87.17$~dBm (Standard Deviation: $14.94$~dBm), and the average SINR is $8.62$~dB (Standard Deviation: $9.67$~dB).}
    \label{fig:traceset}
\end{figure}

The dataset used in the evaluation is the ``LTE-4G-HIGHWAY-DRIVE-TESTS-SALZBURG'' dataset, which was provided by Salzburg Research Forschungsgesellschaft (SRFG)~\cite{6gc4-y070-22}. It is a dataset recorded in real-world environments in Austria, which captures the communication quality of mobile devices.
This dataset comprises active 4G LTE measurements obtained through repeated vehicle-based driving tests conducted on typical Austrian highways from 2018 to 2019 over a period of two years. 
It contains a total of 267,198 measurement data points, including communication metrics such as signal level, RSSI, Signal-to-Interference-plus-Noise Ratio (SINR), RSRP, Reference Signal Received Quality (RSRQ), and instantaneous data rates recorded every second.
It was created to provide researchers with comprehensive and detailed resources for analyzing and studying 4G LTE mobile communication in real-world environments. Sample traces of RSRP and SINR included in the dataset are shown in Fig.~\ref{fig:traceset}.
This dataset records a series of movements covering approximately 25~km from the starting point to the endpoint, grouped into sessions. 
In this study, we use RSRP and SINR from this dataset.

\subsection{Baselines}\label{ssec:Baselines}
To evaluate the estimation accuracy of the proposed LSTM-based LQE model using the datasets in a real-world environment, we compared the performance of the LSTM-based LQE model with that of a baseline model: LQE-SAE~\cite{luo2019link}.
The LQE-SAE model is a state-of-the-art LQE model that employs the latest machine learning techniques. It is known to be one of the most accurate link quality estimation models among existing LQE methods, making it a suitable baseline for our evaluation. 
The LQE-SAE model is built upon an SAE architecture, which considers RSRP and SINR as two separate features. These features are processed by SAEs for both the uplink and downlink, compressing them. 
The compressed features from the two SAEs are further processed by another SAE layer, and then Support Vector Classification (SVC) is used for multi-class classification to estimate link quality grades.
In our evaluation, we use the LQE-SAE model based on the two features, RSRP and SINR. We follow the hyperparameters specified in the paper~\cite{luo2019link} for our evaluation.

\subsection{Parameter Settings}\label{ssec:Settings}
First, the dataset is divided into training, validation, and test datasets in a 7:2:1 ratio.
The validation dataset is used to select the optimal pretrained model, while the test dataset is used to calculate metrics such as accuracy and macro-F1 for the performance evaluation in this section.

The hyperparameters for the preprocess and the LSTM-based LQE model are set as follows.
For the preprocessing, the EMA's span coefficient $\tau$ is set to 120, and the sliding window's window size $N$ is set to 370. 
In the LSTM component, we use a two-layer LSTM with 128 units in the hidden layer. The optimization function is Adam with a learning rate of 0.001. The batch size was set to 128, and the number of epochs was set to 1000. A dropout rate of 0.266 is applied.
For LSTM training, early stopping is enabled with a patience of 50, and the delta parameter is set to $-0.0001$, determining the stopping criterion for training.
In the binning process component, the bin widths are set as shown in TABLE~\ref{tab:rsrp}.

The hyperparameters for the LQE-SAE model are configured as follows:
The LQE-SAE model follows the hyperparameter values used in the paper~\cite{luo2019link}. However, since the first-stage SAE in the LQE-SAE architecture performs feature compression only for RSRP and SINR , we use two SAEs instead of three SAEs.
Additionally, because the dataset we used only contains communication metrics for downlink information, any missing uplink information is zero-padded based on the approach described in Section~\ref{ssec:Preprocess}.

\subsection{Performance Metrics}\label{ssec:PerformanceMetrics}
The performance evaluation of the LQE models was conducted using two multi-class classification evaluation metrics: accuracy and macro-F1.
Due to the uneven distribution of classes within the dataset, accuracy and macro-F1 are used together to provide a more comprehensive evaluation of the model's performance. 

Accuracy measures the ratio of correctly classified samples and is defined by the following equation:
\begin{equation}\label{eq:accuracy}
    Accuracy = \frac{\mathrm{TP} + \mathrm{TN}}{\mathrm{TP} + \mathrm{TN} + \mathrm{FP} + \mathrm{FN}},
\end{equation}
where, $\mathrm{TP}$, $\mathrm{TN}$, $\mathrm{FP}$, and $\mathrm{FN}$ represent the numbers of samples of True Positives, True Negatives, False Positives, and False Negatives, respectively. 
Accuracy quantifies the model's ability to predict correctly across all classes, treating all classes equally.

Macro-F1, on the other hand, calculates the average F1 score across all classes, normalized by the total number of classes $n$, as expressed by the followings:
\begin{equation}\label{eq:macro-f1}
    \text{macro-F1} = \frac{1}{n}\sum_{i=1}^{n}\mathrm{F1}_{i},
\end{equation}
where, $\mathrm{F1}_{i}$ represents the F1 score for the $i$-th class. Macro-F1 is useful when each class is equally important because it evaluates overall performance without disregarding the performance of individual classes.

\subsection{Performance Evaluation}\label{ssec:PerformanceEvaluation}

In this section, we will compare and evaluate the performance of the two models using the accuracy and macro-F1 metrics defined in Section~\ref{ssec:PerformanceMetrics}.
An overview of the estimation results for the SRFG dataset is presented in TABLE~\ref{tab:performance}. In the table, the values highlighted in bold represent the best performance between the two models. 
From TABLE~\ref{tab:performance}, it is evident that the LSTM-based LQE outperforms the traditional LQE-SAE in both accuracy and macro-F1. 
LSTM-based LQE exhibits the highest performance, with accuracy and macro-F1 being 4.0\% and 4.6\% higher, respectively, than LQE-SAE.

\begin{table}[tbp]
    \centering
    \caption{Performance of the LQE-SAE nad LSTM-based LQE models}
    \begin{tabular}{l|l|l}
        \hline
        Model      & Accuracy[\%] & Macro-F1[\%] \\ \hline\hline
        LQE-SAE    & 80.5    & 80.6    \\
        LSTM-based LQE & \textbf{84.5}    & \textbf{85.2}    \\ \hline
    \end{tabular}
    \label{tab:performance}
\end{table}

\section{Conclusions}\label{sec:Conclusion}
In this study, we have developed an LSTM-based WLQE model that can effectively use sequential information. 
We conducted a comparative evaluation with conventional methods using datasets collected from real mobile communication devices, demonstrating that our proposed method enables more accurate estimations.
The results presented in this study highlight the effectiveness of the model using LSTM to estimate RSRP values. This signifies the importance of leveraging temporal correlations in link quality for LQE. 
Moreover, we verified the effectiveness of our approach in real-world environments by training and predicting using datasets containing communication metrics measured by mobile communication devices.

From the experimental results presented in this study, we can envision several promising directions for LSTM-based LQE. 
It can provide crucial information for mobile communication devices using wireless networks to select the optimal link paths and manage data buffering in applications.
In addition, further improvements in accuracy can be achieved by incorporating additional features such as communication metrics not used in this paper, refining discretization methods, and tuning hyperparameter settings.

\section*{Acknowledgments}
This work was partly supported by JSPS KAKENHI Grant Number JP23H03379.

\bibliographystyle{IEEEtran}
\bibliography{IEEEabrv,book_bib}

\end{document}